\documentclass[runningheads]{llncs}
\usepackage[T1]{fontenc}
\usepackage{enumitem}
\usepackage{comment}
\usepackage{siunitx}
\usepackage{tcolorbox}

\usepackage{graphicx}
\begin{document}
\title{An Empirical Study of Multi-Agent RAG for Real-World University Admissions Counseling}
%
%
\author{Anh Nguyen-Duc\inst{1,2} \and
Chien Vu Manh\inst{2} \and
Bao Anh Tran\inst{2} \and Viet Phuong Ngo \inst{2} \and Luan Le Chi\inst{2} \and Anh Quang Nguyen\inst{3}
}
\authorrunning{Anh Nguyen-Duc et al.}
%
\institute{University of South-Eastern Norway, Bo I Telemark, Norway \\
\email{angu@usn.no}
\and
University of Transport Technology, Hanoi, Vietnam 
\\
\email{vumanhchien101@gmail.com,anhtranms03@gmail.com,phuong.nv147@gmail.com,luanlc@utt.edu.vn}
\and
University of Transport and Communications, Hanoi, Vietnam
\\
\email{anhnq@utc.edu.vn}
}
\maketitle              
\begin{abstract}
This paper presents MARAUS (Multi-Agent and Retrieval-Augmented University Admission System), a real-world deployment of a conversational AI platform for higher education admissions counseling in Vietnam. While large language models (LLMs) offer potential for automating advisory tasks, most existing solutions remain limited to prototypes or synthetic benchmarks. MARAUS addresses this gap by combining hybrid retrieval, multi-agent orchestration, and LLM-based generation into a system tailored for real-world university admissions. In collaboration with the University of Transport Technology (UTT) in Hanoi, we conducted a two-phase study involving technical development and real-world evaluation. MARAUS processed over 6,000 actual user interactions, spanning six categories of queries. Results show substantial improvements over LLM-only baselines: on average 92\% accuracy, hallucination rates reduced from 15\% to 1.45\%, and average response times below 4 seconds. The system operated cost-effectively, with a two-week deployment cost of 11.58 USD using GPT-4o mini. This work provides actionable insights for the deployment of agentic RAG systems in low-resource educational settings.

\keywords{RAG  \and Multi Agent Systems \and hybrid RAG \and admission counseling}
\end{abstract}
\section{Introduction}
\label{sec:intro}

The rapid advancement of generative AI (GenAI), especially Large Language Models (LLMs), is transforming multiple sectors—ranging from healthcare to finance and public administration. The last years have witnessed an unprecedented proliferation of GenAI models with the potential to transform educational practice \cite{su__unlocking_2023,mittal_comprehensive_2024,alasadi_generative_2023,li_retrieval-augmented_2025}.
Automated feedback, personalised tutoring, content generation, and conversational search are now technically feasible at scale \cite{alasadi_generative_2023}.
Yet the empirical literature has not kept pace with the release cadence of new foundation models: most claims of effectiveness rest on small demonstrations or synthetic benchmarks rather than controlled studies with authentic educational tasks. Empirical can fill this gap by providing \emph{evidence-based design knowledge}—replicable artefacts, datasets, and evaluations—that inform both practitioners and future model builders.

Admission tasks at higher education is a manual-intensive activity with thousands of high-stakes enquiries into a short time window. Applicants and parents ask about cut-off scores, quotas, scholarship criteria, or the impact of special policies such as regional bonuses and ethnic-minority incentives.
A typical composite question looks like: \emph{“I have 23 transcript points, I belong to Region 1 (mountainous) and priority group 2. How many points do I actually have for admission?”}. Answering correctly requires chaining multiple institutional rules, year-specific cut-off tables, and applicant metadata—not merely general knowledge. Manual hotlines and email desks are slow and error-prone; web FAQs cannot cover the long-tail of personalised scenarios. LLMs could, in principle, deliver instant, dialogic explanations, but off-the-shelf systems (e.g., public ChatGPT or Claude) tend to hallucinate, confuse policy versions, or overlook language-specific details, especially in low-resource languages such as Vietnamese or Thai.

Retrieval-Augmented Generation (RAG) pipelines reduce hallucinations in LLMs by grounding responses in externally retrieved knowledge~\cite{lewis_retrieval-augmented_2020}. Recent advancements enhance this architecture through cross-encoder re-ranking, graph-structured retrieval, and dynamic prompt optimization~\cite{mavromatis_gnn-rag_2024,mishra_searchd_2024}. Building on these foundations, multi-agent LLM systems are emerging as a promising direction for complex reasoning tasks such as university admission counseling. In such systems, distinct agents can specialize in subtasks—e.g., retrieval refinement, regulatory verification, or user context modeling—while coordinating via shared memory or structured prompts to produce coherent answers.

Despite their potential, several open challenges remain. First, domain specificity is a critical concern: it is unclear whether hybrid retrieval combined with multi-agent coordination can consistently extract the correct institutional policies or regulatory clauses required to address nuanced, context-sensitive queries. Second, language representation remains a bottleneck, particularly for low-resource settings. It is uncertain whether Vietnamese-specific or multilingual embedding models offer sufficient semantic fidelity compared to their English-centric counterparts. Finally, prompt engineering strategies—such as few-shot exemplars, chain-of-thought reasoning, or delegated verifier agents—require further empirical validation to assess their effectiveness in minimizing hallucinations while preserving concise, trustworthy outputs in real-world deployments.
 
In this paper, we present MARAUS, a lightweight, domain-specific QA platform tailored for university admission scenarios. Unlike general-purpose LLMs such as ChatGPT, MARUAS can handle specialized queries that require domain logic, score calculations, and local data access. Our experimental results demonstrate that MARUAS outperforms existing LLM-based QA systems in both precision and hallucination control. Furthermore, MARUAS has been deployed in a real-world university admission setting, validating its practical applicability.
\section{Background}
\label{sec:background}

This section reviews key concepts underpinning the design of our system, focusing on recent advances in LLM-based Q\&A educational systems and recent RAG techniques.

\subsection{Educational Q\&A systems}
In the admissions context, RAG-based systems have been developed to manage diverse institutional queries, enhancing both the precision and responsiveness of university application support. For instance, one implementation that combines GPT-3.5 with LlamaIndex has demonstrated superior performance over conventional FAQ systems, particularly in addressing complex student inquiries \cite{chen_facilitating_2024}. Similarly, HICON AI adopts a RAG-based approach to deliver tailored college admissions advice by segmenting applicants into advisory profiles and incorporating resume analysis for individualized suggestions \cite{singla_hicon_2024}.
In Vietnamese admission context, Bui et al. consolidate educational FAQs, curricular resources, and data from learning management systems (LMS) into a cohesive knowledge graph \cite{bui_cross-data_2024}. This graph-driven design enhances both intent detection and policy alignment. As shown in Table 1, our proposed approach MARAUS adopting novel strategies basing on current available GPT models with superior performance comparing to existing studies.

\begin{table}[ht]
\centering
\resizebox{\textwidth}{!}{%
\begin{tabular}{|p{4.2cm}|p{3.2cm}|p{3.2cm}|p{4.4cm}|}
\hline
\textbf{Reference} & \textbf{Architecture} & \textbf{Retrieval} & \textbf{LLM Model} \\
\hline
Bui et al. 2024 \cite{bui_cross-data_2024} & Knowledge Graph & Semantic Similarity & URA \\
\hline
Singla et al. 2024 \cite{singla_hicon_2024} & Singular RAG & Semantic Similarity & LlaMA2 \\
\hline
Ehrenthal et al., 2024 & Unknown & Singular RAG & PaLM~2 for Text \\
\hline
Z.\ Chen et al. 2024 \cite{chen_facilitating_2024} & Singular RAG & Semantic Similarity & Text-davinci-003, GPT-3.5-Turbo \\
\hline
This study & Multi-agents & Keyword, semantic and Re-rank with LLM & GPT-4o mini \\
\hline
\end{tabular}}
\caption{Overview of Retrieval-Augmented Generation configurations used in recent studies.}
\label{tab:rag-configs}
\end{table}

\subsection{Retrieval Strategies for RAG Systems}
\label{sec:retrieval-strategies}

Retrieval forms the backbone of RAG systems, providing the contextual information that language models use to generate relevant and accurate responses. Broadly, retrieval approaches fall into three categories: keyword-based, semantic vector-based, and hybrid methods.

Keyword-based retrieval, such as BM25 \cite{robertson_probabilistic_2009}, ranks documents based on term frequency and exact matches. These methods are efficient, interpretable, and widely used in production through engines like Lucene and ElasticSearch. However, they often fail when queries use synonyms, informal language, or region-specific phrasing—as is common in Vietnamese higher education contexts.

Semantic retrieval methods address these limitations by embedding text into dense vector spaces using pretrained models like SBERT \cite{reimers_sentence-bert_2019} and MPNet \cite{song_mpnet_2020}. Tools like FAISS perform fast approximate nearest-neighbor search in this space, enabling robust retrieval for paraphrased or imprecise queries. Nonetheless, semantic similarity alone can retrieve passages that are topically related but contextually irrelevant.

To balance precision and recall, hybrid retrieval strategies combine both lexical and semantic signals. In our system, top-$k$ results are retrieved using both FAISS and BM25, then re-ranked using a GPT cross-encoder. This cross-encoder jointly processes the query and each candidate passage to assign relevance scores, improving the accuracy of context selection. This approach is particularly effective in disambiguating overlapping programme names or policy terms, and ensures that retrieved content is both semantically aligned and task-relevant.

\section{Research Approach}
\subsection{Research Design}
\label{sec:design}
We employ an \emph{in‐depth single–case study} design~\cite{runeson_guidelines_2008}. Case studies are regarded as the most suitable empirical strategy when (i) the phenomenon under investigation cannot be separated from its real‐world context, and (ii) the goal is to build rich, explanatory theory rather than derive statistical generalisations. Our objective is to understand \emph{how} and \emph{to what extent} RAG grounded in Large Language Models (LLMs) can support real-world admission activities. Data collection was between Jan 2025 and July 2025 and conducted in two phases:
\begin{itemize}
    \item Phase 1 - Technological experimentation: the admission workflow was explored, requirements were collected, datasets were collected, models were experimented with and technologically optimized.
    \item Phase 2 - Process experimentation: A prototype of AI conversational system was built and adopted by real users. Evaluation metrics were collected in 2 weeks.
\end{itemize}

\subsection{Our case}
\textbf{University of Transport Technology (UTT)} in Hanoi, Vietnam, has several characteristics for our case study. The university handles over \num{10,000} applicant enquiries annually across peak periods, typical for medium-sized public universities in Vietnam. They have large volume of digital conversations - via three channels, Facebook, Zalo and Direct contact, there are more than 4.5 GB of conversational data since 2022. Most importantly, UTT’s senior leadership committed to providing sustained access to operational logs, historical queries, and key admission personnel throughout the design and evaluation phases.

The existing enquiry-handling process at UTT is decentralised, manually intensive, and distributed across multiple communication channels. It unfolds as follows: (1) static publication, admission information are first published on the university’s website, Facebook page, and printed leaflets. This information is updated annually, often within tight timelines dictated by the Ministry of Education and Training (MOET), (2) query intake. Prospective students and parents submit questions through five main channels: (i) telephone hotline, (ii) university e-mail, (iii) Facebook Messenger, (iv) on-campus consultation booths, and (v) informal messaging platforms (e.g., Zalo, SMS), (3) Manual triage and assignment. A limited number of admission officers—typically fewer than 10 during peak weeks—monitor all incoming messages. Questions are manually categorised, sometimes assigned to relevant departments (e.g., financial aid), and queued for response. Duplication is high: some FAQs (e.g., cut-off scores, tuition fees) appear hundreds of times in a single day. (4) - Knowledge transfer, officers consult a shared internal Google Drive folder or archived email chains for reference documents, then draft individual responses. In practice, replies are often copy-pasted or rephrased. However, inconsistencies frequently arise due to non-synchronised document versions or incomplete updates.

Despite genuine efforts by staff, the current workflow suffers from three structural bottlenecks: (1) high-volume repetition of FAQ-type queries, (2) lack of integrated knowledge management, and (3) inability to personalise responses to individual applicants. These factors make UTT an ideal site to assess whether RAG systems can meaningfully improve both the accuracy and responsiveness of admission counselling under realistic operational constraints.

\subsection{Threats to Validity}
\label{sec:validity}

As with any single-case empirical study, several threats to validity must be considered~\cite{runeson_guidelines_2008}. One key threat lies in how chatbot performance is operationalised. We provided standard experimental metrics, i.e. accuracy, response time and user rate with real-user data by recording the number of correct and incorrect responses observed during actual interactions. Another limitation concerns the generalisability of the findings, as the study was conducted within a single institution. Institutional practices related to admissions, data governance, and IT infrastructure vary considerably across contexts. However, UTT’s characteristics—a mid-sized public university with decentralised communication and limited infrastructure—are broadly representative of many Vietnamese and Asian/ African higher education institutions. We describe UTT’s specific workflow in Section 3.2. to support analytical generalization. Reproducibility also presents a potential threat. To mitigate this, we have documented all configuration parameters in detail and publicly released our codebase and test dataset. This enables other researchers and practitioners to replicate or adapt our work in similar environments.

\section{MARAUS development}
This section presents the design, implementation, and experimental evaluation of MARAUS (Multi-Agent and Retrieval Augmented University Admission System), a hybrid Retrieval-Augmented Generation (RAG) platform for handling diverse user queries in university admission contexts. MARAUS integrates multi-agent orchestration, semantic retrieval, hybrid re-ranking, and large language model (LLM) generation, with a focus on precision, interpretability, and hallucination control.
\subsection{System architecture}
\begin{figure}[ht]
  \centering
  \includegraphics[width=\textwidth]{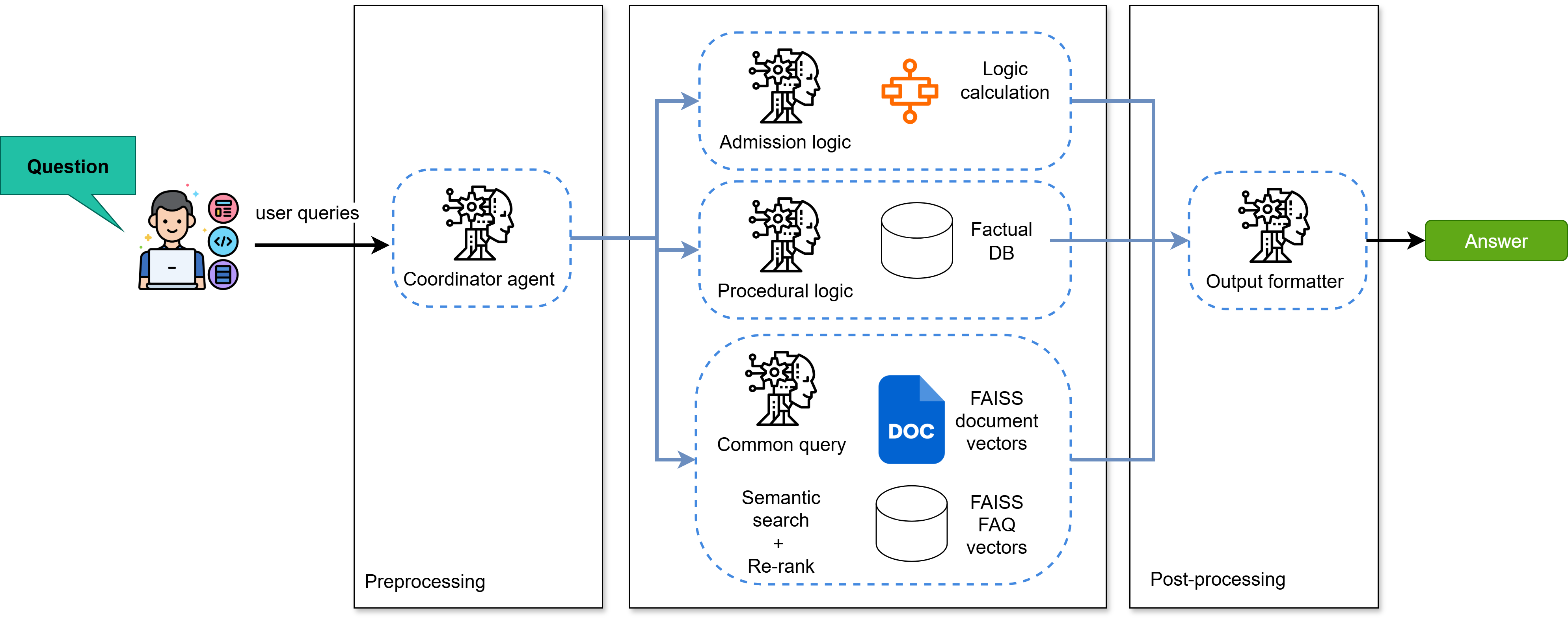}
  \caption{An overview of the MARAUS system}
  \label{fig:multiagent-rag}
\end{figure}

The core of MARAUS is a multi-agent coordinator that classifies incoming queries into four distinct processing pipelines (Figure~\ref{fig:multiagent-rag}):
\begin{itemize}
    \item Information search agent: Executes keyword-based and semantic retrieval for informational queries (e.g., program details).
    \item Score calculation agent: Handles numeric computation for score-related queries, extracting structured attributes to compute transcript points, priority bonuses, and total eligibility scores.
    \item Recommendation agent: for queries predicting program eligibility, such as “With XX transcript points, can I pass [program]?” or “What programs can I pass with XX exam points?”, the system extracts information
    \item General Query agent: Applies a fallback hybrid RAG strategy when query classification confidence is low
\end{itemize}

\subsection{Pre-processing}
All textual data undergoes rigorous preprocessing. We remove boilerplate content, HTML artifacts, and near-duplicates using a Jaccard similarity filter ($<0.9$). Text normalization includes lowercasing, diacritic normalization, and tokenization via \textsf{VnCoreNLP}. Personally identifiable information (PII) is redacted using regular expression patterns targeting phone numbers, emails, and national IDs. Additionally, entropy-based filters detect and exclude nonsensical input or prompt-injection attempts.

Document corpora are segmented into \num{8,412} overlapping context windows (500-token size, 100-token stride), embedded with the \texttt{XAI-encode-all-mpnet-base-v2} model (\num{768} dimensions, \SI{28}{ms}/chunk on CPU). The resulting vectors are stored in a FAISS \emph{IndexFlatIP}, occupying \SI{565}{MB} of RAM. Full index construction completes in \SI{41}{s} on an 8-core machine.

The training set includes \textbf{1,376 FAQ pairs} and \textbf{98 document chunks} (approx.\ 600–700 tokens per chunk), curated from UTT’s internal systems (MongoDB, Google Drive). Chunk size was empirically tuned to balance context retention against LLM token constraints.

\subsection{Hybrid Retrieval in MARAUS} We employ a hybrid RAG pipeline, combining dense vector retrieval and sparse keyword search.

\textbf{Semantic Retrieval}. At runtime, user queries are embedded into a \num{768}-dimensional vector using \texttt{Xenova/all-mpnet-base-v2}. FAISS retrieves the top-$k$ (k=15) similar FAQ entries and the top document chunk based on cosine similarity. A filtering threshold of 0.9 ensures high semantic relevance.

\textbf{Keyword Retrieval and Fusion}. To handle spelling variations and rare keyword queries, an ElasticSearch 8.11 BM25 index runs in parallel. The union of BM25 and FAISS outputs is passed to a hybrid re-ranking stage.

\textbf{Re-Ranking with LLM}. We deploy GPT-4o mini as a zero-shot cross-encoder for relevance scoring. Each candidate passage is concatenated with the user query in a structured prompt, and GPT-4o mini assigns a normalized relevance score [0,1]. The top-2 passages are retained. This step reduces false positives by \SI{38}{\%} compared to raw FAISS retrieval.

\subsection{Post processing}
The final prompt comprises of (1) an instruction block discouraging speculation, (2) the two top-ranked passages, explicitly cited by ID, and (3) the user query. Output text generation is performed using GPT-4o with \texttt{temperature=0.7}, \texttt{top p=0.9}, and a \texttt{max tokens=350} limit. A streaming interface returns the first token within \SI{350}{ms} on average. \textbf{Hallucination Mitigation}. A custom post-processor enforces citation integrity. If generated answers lack at least one passage citation, they are discarded and regenerated with penalized decoding parameters. This mechanism reduces hallucination rates from \SI{15}{\%} (LLM-only) to \SI{1.45}{\%} in the hybrid pipeline.

\subsection{Experimental Evaluation}
We evaluate MARAUS across multiple configurations: LLM-only, RAG+Re-rank, and Hybrid RAG. Metrics include precision, recall, F1-score, response time, and hallucination rate. All experiments were conducted on an 8-core Intel Xeon machine with \SI{32}{GB} RAM. The experiments were on a set of 100 pairs of questions and answers with grounded truth provided by UTT. As shown in Table 2, the Hybrid RAG pipeline outperforms baselines across all metrics, achieving near-perfect precision and significantly reducing hallucination rates, while maintaining sub-4-second response times. These results demonstrate the effectiveness of multi-agent coordination, hybrid retrieval, and LLM-controlled re-ranking in domain-specific QA systems, helping us to define the retrieval approach for MARAUS

\begin{table}[ht]
\centering
\caption{MARAUS Experimental Results Summary}
\label{tab:exp-summary}
\begin{tabular}{|l|c|c|c|}
\hline
\textbf{Metric} & \textbf{LLM Only} & \textbf{RAG+Re-rank} & \textbf{Hybrid RAG} \\
\hline
Precision & 0.70 & 0.90 & \textbf{0.985} \\
Recall & 0.65 & 0.85 & \textbf{0.89} \\
F1-score & 0.67 & 0.87 & \textbf{0.91} \\
Response Time (s) & 7.0 & 4.0 & \textbf{3.75} \\
Hallucination Rate & 15\% & 6\% & \textbf{1.45\%} \\
\hline
\end{tabular}
\end{table}

\begin{tcolorbox}[title=Key Observations,colback=gray!5,colframe=black!75,boxrule=0.8pt]
\textbf{KO1} – Multi-agent RAG pipelines can significantly enhance the reliability and factual grounding of LLMs in complex, real-world advisory tasks, reducing hallucination while maintaining high accuracy.
\end{tcolorbox}

\section{Result}
\label{sec:case-eval}
This section presents the data extracted from two weeks publishing MARAUS, evaluation metrics and our findings.
\subsection{Data}
Data was collected from 2-week running of the Q \& A system for UTT during summer 2025. There are typically six types of user questions, reflecting increasing levels of complexity and interaction. The first type is Simple Keyword Retrieval, where the system extracts direct answers from FAQs or databases based on keywords, such as "Does the university offer medical programs?". The second type involves Intent and Entity Recognition, where the chatbot identifies the category of the question and extracts key details like majors, scores, or locations, for example: "Is 25 points enough for Computer Science?". The third type is Answer Generation, where the chatbot generates new responses for questions not directly covered in the knowledge base, such as: "Give me a brief introduction to Artificial Intelligence." The fourth type is Logical Reasoning and Calculation, where the system performs tasks like calculating admission scores or applying priority points, for example: "I have 25.5 points from my high school record, in a KV1 area, can I get into Computer Science?" The fifth type is Personalization and Multi-turn Conversation, where the chatbot maintains context across multiple turns, such as remembering a user mentioned they are from Quang Tri in one message and responding appropriately to follow-up questions about priority scoring. Finally, the sixth type is Handling Ambiguous or Subjective Questions, which requires the chatbot to analyze user sentiment, social context, and implied goals, such as: "Which major is suitable for introverts?". The total of 6079 pairs of Questions and Answers were manually inspected by coauthors of this paper. 
 
\subsection{Evaluation metrics}
We used the following metrics to evaluate MARAUS:
\begin{itemize}
    \item Number of questions (items) per day: Each item refers to a pair of question and answer extracted from the MARAUS chat logs.
    \item Number of tokens per day: The total number of input and output tokens recorded from the OpenAI platform.
    \item Accuracy: The ratio of correct answers to the total number of questions. A correct answer is defined as one that provides neither incorrect information nor irrelevant responses to user questions.
    \item Perceived user satisfaction: The rating given by university staff members regarding the chatbot's performance.
\end{itemize}

\subsection{Findings}

Number of token used ranges from 350k to 1.4mil token every day. With the average input token per message is 1748 tokens and average output token per message is 60 tokens.
The cost for two weeks actual running of the bot is 11.58 USD with GPT-4o mini or 57.90 USD with GPT-4o.


As shown in Figure 2, the analysis of MARAUS chatbot interactions over the observed period shows a high proportion of correct answers across all days, with daily accuracy consistently ranging from 87\% to 94\%. The total number of questions per day varies significantly, from as few as 180 interactions on 05/07 to as many as 797 interactions on 24/06. On average the accuracy is 92\%. The distribution of wrong answers are different across categories of questions relating Personalization and Multi-turn conversation and Handling ambiguous and subjective questions. There are also many wrong answers due to a misunderstanding of the questions by users when the use of language is vague or inaccurate.

\begin{figure}[ht]
  \centering
  \includegraphics[width=\textwidth]{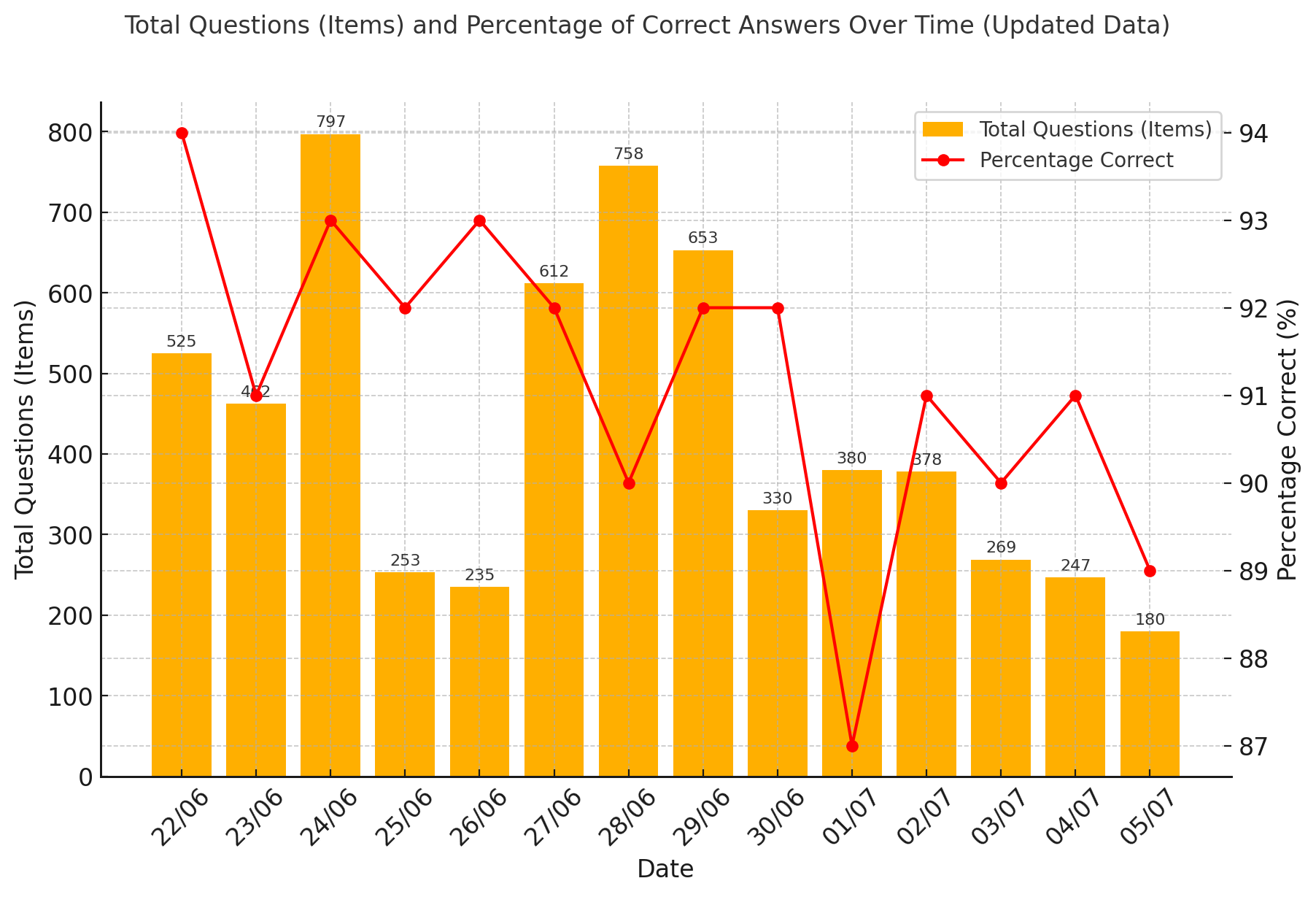}
  \caption{Systems' accuracy}
  \label{fig:dailytoken}
\end{figure}

User satisfaction averaged 4.5/5 across 7 admission officers and test users. Officers reported a significant reduction in repetitive question handling and favoured the explainability provided by chunk citations. The officers also appreciated the personalised responses tailored to score ranges and programm preferences.

\begin{tcolorbox}[title=Key Observations,colback=gray!5,colframe=black!75,boxrule=0.8pt]
\textbf{KO2} – Domain-specialized orchestration that integrates retrieval, reasoning, and personalized interaction is essential for addressing diverse user intents beyond standard FAQ matching in practical deployments
\end{tcolorbox}
\section{Discussion}
Recent studies have explored various approaches to enhance the accuracy and relevance of AI chatbots using RAG techniques combined with large language models (LLMs). For instance, Singla et al. developed the HiCON chatbot, integrating Meta Llama 2 with RAG to handle four types of user questions \cite{singla_hicon_2024}. Their system achieved an average accuracy of 80\%, demonstrating the potential of combining LLMs with retrieval pipelines to improve response quality. Similarly, Chen et al. proposed a system based on ChatGPT-3.5 with RAG using LlamaIndex \cite{chen_facilitating_2024}. Their approach significantly improved the baseline performance, increasing the average accuracy from 41.4\% (ChatGPT-3.5 alone) to 89.5\% with their RAG-enhanced chatbot, and achieving a peak accuracy of 94.7\% in specific scenarios.

Within the Vietnamese context, Bui et al. explored a knowledge graph-based approach for question answering in the education domain. While promising in concept, their work remains at a preliminary stage, with limited experimental data and no large-scale deployment or stress testing reported.In contrast, our proposed system offers a cost-effective and scalable solution. By leveraging multi-agent strategies and off-the-shelf LLM components combined with RAG, our approach enables feasible real-world adoption without the overhead of maintaining complex knowledge graphs. Furthermore, the system is tested with a larger and more diverse dataset, covering six categories of user questions—including reasoning, personalization, and handling subjective queries—thereby increasing its applicability in real-world university counselling services.

\section{Conclusion}
This study presents MARAUS, an operational, multi-agent RAG-based chatbot designed for university admissions counseling. Unlike prior studies with small-scale validation, MARAUS was tested in a real-world deployment, engaging directly with thousands of applicants and university staff over an intensive admissions period. Our findings show that MARAUS outperforms traditional LLM-only approaches and prior RAG-based systems. The integration of hybrid retrieval, LLM-based re-ranking, and multi-agent task specialization contributed significantly to these outcomes. The study also highlights practical considerations for deployment in low-resource language settings, where domain-specific knowledge, policy variations, and language nuances pose additional challenges. 

Future work will extend MARAUS to other universities and explore broader domains such as financial aid advising or career counseling. We also plan to refine the system’s personalization capabilities and investigate adaptive learning from continuous user feedback to further enhance long-term performance.

\bibliographystyle{splncs04}
\bibliography{main.bib}

\end{document}